\documentclass[12pt]{amsart}
\usepackage[russian,english]{babel}
\usepackage{amsfonts,amssymb,amsthm}
\usepackage[mathscr]{eucal}
\usepackage{hyperref}
\usepackage[matrix,arrow]{xy}
\usepackage{times}

\title[ Lagrange Anchor for Bargmann-Wigner equations]
 {Lagrange Anchor  for \\ Bargmann-Wigner equations}

\author{D.S. Kaparulin,  S.L.  Lyakhovich and A.A. Sharapov}
\address{Department of Quantum Field Theory, Tomsk State University, Lenin ave. 36, Tomsk 634050, Russia.}

\email{dsc@phys.tsu.ru, sll@phys.tsu.ru, sharapov@phys.tsu.ru}

\begin{document}
\maketitle
\begin{abstract}
A Poincar\'e  invariant Lagrange anchor is found for the
non-Lagrangian relativistic wave equations of Bargmann and Wigner
describing free massless fields of spin $s > 1/2$ in
four-dimensional Minkowski space. By making use of this Lagrange
anchor, we assign a symmetry to each conservation law.
\end{abstract}

\section*{Introduction}

The notions of symmetry and conservation law are of paramount
importance for classical and quantum field theory. For Lagrangian
theories both the notions are tightly connected to each other due to
Noether's first theorem. Beyond the scope of Lagrangian dynamics,
this connection has remained unclear, though many particular results
and generalizations are known (see \cite{K-S} for a review). In our
recent works \cite{KLS2,KLS3} a general method has been proposed for
connecting symmetries and conservation laws in not necessarily
Lagrangian field theories. The key ingredient of the method is the
notion of a Lagrange anchor introduced earlier \cite{KLS} in the
context of quantization of (non-)Lagrangian dynamics. Geometrically,
the Lagrange anchor defines a map from the vector bundle dual to the
bundle of equations of motion to the tangent bundle of the
configuration space of fields such that certain compatibility
conditions are satisfied. The existence of the Lagrange anchor is
much less restrictive for the equations than the requirement to be
Lagrangian or admit an equivalent Lagrangian reformulation.

The theory of massless higher-spin fields is an area of particular
interest for application of the Lagrange anchor construction. Here one can keep
in mind Vasiliev's higher-spin
equations in the form of unfolded representation \cite{V1,V2,BCIV}.
The unfolded field equations are not Lagrangian even at the free level
and their quantization by the conventional methods is
impossible. Finding a Lagrange anchor for these
equations can be considered as an important step towards  the consistent quantum theory of higher-spin fields. In our recent paper \cite{KLS1}, a general construction for the Lagrange anchor was proposed for unfolded equations that admit an equivalent Lagrangian formulation.

In this paper,  the general concept of Lagrange anchor is
exemplified by the Bargmann-Wigner equations for free massless
fields of spin $s\geq 1/2$ in the four-dimensional Minkowski space
\cite{PR}. The choice of the example is not accidental. First of
all, it has long been known that the model admits infinite sets of
symmetries and conservation laws. These have been a subject of
intensive studies by many authors during decades, see e.g.
\cite{Lipkin,Morgan,Kibble,Fairlie,FN,KVZ,AP0,GSV} and references
therein. However, a complete classification has been obtained only
recently, first for the conservation laws \cite{AP1} and then for
the symmetries \cite{AP2}. As the field equations are non-Lagrangian
for $s> 1/2$, there is no immediate Noether's correspondence between
symmetries and conservation laws. The rich structure of symmetries
and conservation laws in the absence of a Lagrangian formulation
makes this theory an appropriate area for testing the concept of
Lagrange anchor.

\section{The Lagrange anchor in field theory}

In this section we give a brief exposition of the Lagrange anchor
construction. A more detailed discussion can be found in \cite{KLS}.

Consider a collection of fields $\phi^i(x)$ whose dynamics are governed by a system of PDEs
\begin{equation}\label{T_a}
T_{a}(x, \phi^i(x), \partial_\mu\phi^i(x), \ldots)=0\,.
\end{equation}
Here  $x$'s denote local coordinates on a space-time manifold $X$
and indices $i$ and $a$ numerate the components of fields and field
equations. As we do not assume the field equations (\ref{T_a}) to
come from the least action principle, the indices $i$ and $a$ may
run through different sets. In what follows we  accept Einstein's
convention on summation by repeated indices.

Instead of working with the set of PDEs (\ref{T_a}) it is convenient
for us to introduce a single linear functional
$$
T[\xi]=\int_X dx \xi^a
T_{a}
$$
of the test functions $\xi^a=\xi^a(x)$ with compact support. Then $\phi^i(x)$ is a solution to (\ref{T_a}) iff $T[\xi]=0$ for all $\xi$'s.

Consider now the  linear space of the variational vector fields of the form
\begin{equation}\label{Vzeta}
{V}[\xi]=\int_X dx
{V}^i(\xi)\frac{\delta}{\delta\phi^i(x)}\,,
\end{equation}
where ${V}^i(\xi)=\hat{V}^i_a\xi^a(x)$ and
\begin{equation*}
\hat{V}^i_a=\sum_{q=0}^p V_a^{i,\mu_1,\ldots,\mu_q}(x, \partial_\mu \phi(x), \ldots)
\partial_{\mu_1}\ldots\partial_{\mu_q}
\end{equation*}
is a matrix differential operators with coefficients being smooth functions of space-time coordinates, fields and their partial derivatives up to some finite order. Action of the variational vector fields on local functionals of $\phi$'s is defined by the usual rules of variational calculus.

The variational vector field (\ref{Vzeta}) is called the
\textit{Lagrange anchor} if for any $\xi_1$ and $\xi_2$ there exist
a test function $\xi_3$ such that the following condition is
satisfied:
\begin{equation}\label{LA}
V[\xi_1]T[\xi_2]-V[\xi_2]T[\xi_1]=T[\xi_3]\,.
\end{equation}
Clearly, if exists, the  function $\xi_3$ is given by a bilinear differential operator
acting on $\xi_1$ and $\xi_2$:
\begin{equation}\label{C}
\xi_3^a={C}^a(\xi_1,\xi_2)\,.
\end{equation}
The coefficients of the operator ${C}$ may depend on space-time
coordinates $x$, fields $\phi$ and their derivatives.

The defining condition (\ref{LA}) means that the left hand side
vanishes whenever $\phi$'s satisfy the field equations (\ref{T_a}).

The Lagrangian equations $\delta S/\delta\phi^i(x)=0$ admit an identical (or canonical)
Lagrange anchor determined by the operator $\hat{V}_i^j=\delta^j_i$.  The defining condition (\ref{LA}) reduces to
commutativity of variational derivatives
$$
\frac{\delta^2 S}{\delta\phi^i(x)\delta\phi^j(x')}=\frac{\delta^2 S}{\delta\phi^j(x')\delta\phi^i(x)}\,.
$$
If the Lagrange anchor
is invertible in the class of differential operators, then the operator $\hat{V}^{-1}$ have the sense of an integrating multiplier in the inverse problem of calculus of variations. In this
case, one can define the local action functional $S[\phi]$ such that
$\delta S/\delta\phi^i=\hat{V}^{-1}_i(T)$.

The classification of  Lagrange anchors for the equations of
evolutionary type was obtained in \cite{KLS-ODE}. In particular, it
was shown that all the stationary and strongly integrable (we
explain the notion of integrability below) Lagrange anchors for
determined systems of evolutionary equations are in one-to-one
correspondence with the Poisson structures that are preserved by
evolution. Let us illustrate this fact by the example of autonomous
system of ODEs in normal form
$$
\dot{y}^i=F^{i}(y)\,.
$$
Consider the following ansatz for the Lagrange anchor:
\begin{equation}\label{a}
V[\xi]=\int dt V^{ij}(y(t))\xi_j(t)\frac{\delta}{\delta y^i(t)}\,.
\end{equation}
Here $V^{ij}(y)$ is a contravariant tensor on the space of $y$'s.  Verification of the
defining condition (\ref{LA}) yields
\begin{equation}\label{alpha}
V^{ij}+V^{ji}=0\,,\qquad F^k{\partial_k V^{ij}}+V^{ik}{\partial_k F^j}-V^{jk}{\partial_k
F^i}=0\,,
\end{equation}
that is, $V^{ij}(y)$ must be an $F$-invariant bivector field on the phase space of the system.
The corresponding bidifferential operator (\ref{C}) is given by
$$
\xi^3_k=\partial_kV^{ij}\xi^1_i\xi^2_j\,.
$$
(In this particular case it does not involve derivatives of $\xi_1$ and $\xi_2$.)

One more important notion related to the Lagrange anchor is that of integrability.
The Lagrange anchor is said to be \textit{strongly integrable} if the following two conditions are satisfied:
\begin{equation}\label{AAn}
\begin{array}{c}
[{V} [\xi_1], {V} [\xi_2]]= {V}[C(\xi_1,\xi_2)]\,,\\[3mm]
C^a(\xi_1,C(\xi_2,\xi_3))+
V[\xi_1]C^a(\xi_2,\xi_3)+cycle(\xi_1,\xi_2,\xi_3)=0\,.
\end{array}
\end{equation}
The first condition means that the variational vector fields $V[\xi]$ form an integrable distribution in the configuration space of fields. If the Lagrange anchor is \textit{injective}, that is, $V[\xi]=0$ implies $\xi=0$, then the second relation follows from the first one due to the Jacobi identity for the commutator of vector fields. Taken together relations (\ref{AAn}) define what is known in mathematics as the Lie algebroid with anchor $V$ and bracket $C$, see e.g. \cite{Mc}.

The canonical Lagrange anchor is strongly integrable since $C=0$ in this case. The integrability condition for (\ref{a}) requires the bivector $V=V^{ij}(y)\partial_i\wedge\partial_j$ to satisfy the Jacobi identity
$$
V^{in}\partial_nV^{jk}+cycle(i,j,k)=0\,.
$$

It should  be noted, that the strong integrability condition is
\textit{not} a part of  the definition of Lagrange anchor. In many
cases it can be considerably relaxed or even omitted. So, in
general, the concept of Lagrange can not be substituted by that of
Lie algebroid. A lot of examples of non-canonical Lagrange anchors
for non-Lagrangian and non-Hamiltonian theories can be found in
\cite{KLS2,KLS,KLS1,LS2,LSDUY,LS3,KLS5}.

\section{The generalization of Noether theorem for non-Lagrangian theories}

A vector field $j^\mu(x,\phi^i,\partial_\mu\phi^i,\ldots)$ on $X$ is
called a conserved current if its divergence is proportional to the
equations of motion (\ref{T_a}),  i.e.,
\begin{equation}\label{dj}
 \partial_\mu j^\mu=\sum_{q=0}^p\Psi^{a,\mu_1\ldots\mu_q}
 (x,\phi^i(x),\partial_{\mu}\phi^i(x),\ldots)\partial_{\mu_1}\ldots\partial_{\mu_q}T_a\,.
\end{equation}
The right hand side is defined by some differential operator  $\Psi$ called the characteristic of the conserved current $j$. Two
conserved currents  $j$ and $j'$ are considered to be  equivalent if
$j^{\mu}-j'^{\mu}=\partial_{\nu}i^{\nu\mu}$ $
({mod\phantom{a}} T_a)$ for some bivector
$i^{\mu\nu}=-i^{\nu\mu}$. Similarly, two characteristics $\Psi$ and
$\Psi'$ are said to be equivalent if they correspond to equivalent
currents. These equivalences can be used to simplify the form of
characteristics. Namely, one can see that in each equivalence class of $j$ there is a representative with
$\Psi$ being the zero order differential operator $\Psi^a$.
For such a representative equation (\ref{dj}) can be written as
\begin{equation}\label{TP}
T[\Psi]=\int_X \partial_\mu j^\mu\,.
\end{equation}
It can be shown that there is a one-to-one correspondence between equivalence
classes of conserved currents and characteristics \cite{KLS2}.

Given a  Lagrange anchor, one can assign to any characteristic
$\Psi$ a variational vector field $V[\Psi]$. The main observation
made in \cite{KLS2} was that $V[\Psi]$ generates a symmetry of the
field equations (\ref{T_a}):
\begin{equation}\label{deltaT}
\delta_\varepsilon \phi^i=\varepsilon V^i(\Psi)\,, \qquad
\delta_\varepsilon T[\xi]=\varepsilon V[\Psi]T[\xi]=\varepsilon
T[C(\Psi,\xi)-V[\xi]\Psi]\,,\end{equation} with $\varepsilon$ being
an infinitesimal constant parameter. These relations  follow
immediately from the definitions of the Lagrange anchor (\ref{LA})
and characteristic (\ref{TP}) upon substitution $\xi_1=\Psi$.

Recall that any characteristic $\Psi$ of Lagrangian equations
$\delta S/\delta \phi^i(x)=0$ generates a symmetry
$\delta_\varepsilon\phi^i=\varepsilon \Psi^i$ of the action
functional and thus  the equations of motion. This statement
constitutes the content of Noether's first theorem \cite{K-S} on
correspondence between symmetries and conservations laws. One the
other hand, this correspondence is a simple consequence of a more
general relation (\ref{deltaT}) if one takes  the canonical Lagrange
anchor $V^i(\xi)=\xi^i$ for Lagrangian equations. From this
perspective, the assignment \begin{equation}\label{NT} \Psi\mapsto
V[\Psi]
\end{equation}
can be regarded as a generalization of the first Noether's theorem to the case of non-Lagrangian PDEs. In general,  the map (\ref{NT})
from the space of characteristics (= conservation laws)  to the space  of symmetries is
neither surjective nor injective. The symmetries from the image of this map
are called  \textit{characteristic symmetries}.

In the particular case of strongly integrable Lagrange anchor the space of characteristics can be endowed with the structure of Lie algebra. The corresponding Lie bracket reads
\begin{equation}\label{DB}
\{\Psi_1,\Psi_2\}^a=V[\Psi_1]\Psi_2^a-V[\Psi_2]\Psi^a_1+C^a(\Psi_1,\Psi_2)\,.
\end{equation}
Furthermore, the anchor map (\ref{NT}) defines a homomorphism from the Lie algebra of characteristics to the Lie algebra of symmetries
$$
[V[\Psi_1], V[\Psi_2]]=V[\{\Psi_1,\Psi_2\}]\,.
$$
The bracket (\ref{DB}) generalizes  the Dickey bracket of conserved
currents \cite{D} known in Lagrangian  dynamics.

\section{The Lagrange anchor and characteristic symmetries for
the Bargmann-Wigner equations}

In this section we illustrate the general concept of Lagrange anchor
by the example of Bargmann-Wigner's equations. These equations
describe free massless fields of spin $s>0$ on $d=4$ Minkowski
space. The equations read
\begin{equation*}
T^{\dot\alpha}_{\alpha_1\cdots\alpha_{2s-1}}:=\partial^{\alpha\dot{\alpha}}\varphi_{\alpha\alpha_1\ldots\alpha_{2s-1}}=0\,,
\end{equation*}
where $\varphi_{\alpha_1\ldots\alpha_{2s}}(x)$ is a symmetric,
complex-valued spin-tensor field on $\mathbb{R}^{3,1}$. We use the
standard notation of the two-component spinor formalism \cite{PR},
e.g.
$\partial^{\alpha\dot{\alpha}}=(\sigma^\mu)^{\alpha\dot{\alpha}}\partial/\partial
x^{\mu}$, $\mu=0,1,2,3$, $\alpha,\dot\alpha=1,2$, and the spinor
indices are rised/lowered with $\varepsilon_{\alpha\beta}$,
$\varepsilon_{\dot\alpha\dot\beta}$ and the inverse
$\varepsilon^{\alpha\beta}$, $\varepsilon^{\dot\alpha\dot\beta}$.

To make contact with the general definitions of the previous section
let us mention that the indices of equations and fields are given by the multi-indices
$a=(\dot\alpha,\alpha_1,\ldots,\alpha_{2s-1})$ and
$i=(\alpha_1,\ldots,\alpha_{2s})$. It is well known that the Bargmann-Wigner equations are non-Lagrangian unless $s=1/2$.

In \cite{KLS5}, it was shown that the Bargmann-Wigner equations admit the following  Pioncar\'e-invariant and strongly
integrable Lagrange anchor:
\begin{equation}\label{ULA}
V(\xi)_{\alpha_1\cdots\alpha_{2s}}=i^{2s}\partial_{(\alpha_2\dot\alpha_2}\cdots\partial_{\alpha_{2s}\dot\alpha_{2s}}
 \bar{\xi}_{\alpha_1)}^{\dot\alpha_2\cdots\dot\alpha_{2s}}\,.
\end{equation}
The round brackets mean  symmetrization. This Lagrange anchor is
unique (up to equivalence) if the requirements of (i)
field-independence, (ii) Pioncar\'e-invariance and (iii) locality
are imposed. Being independent of fields, the Lagrange anchor is
integrable with $C=0$.

Let $\Psi$ be a characteristic of a conserved current $j$ such that
$$\partial^{\alpha\dot\alpha}j_{\alpha\dot\alpha}=\Psi_{\dot\alpha}^{\phantom{\dot\alpha}\alpha_1\ldots\alpha_{2s-1}}
T^{\dot\alpha}_{\phantom{\dot\alpha}\alpha_1\ldots\alpha_{2s-1}}+ c.c.\,.$$
Then the Lagrange anchor (\ref{AAn}) takes this characteristic
to the symmetry
\begin{equation}\label{CSym}
\delta_\varepsilon \varphi_{\alpha_1\ldots\alpha_{2s}}=\varepsilon V(\Psi)_{\alpha_1\cdots\alpha_{2s}}\,,
\end{equation}
where $V(\Psi)$ is defined by (\ref{ULA}).

Applying (\ref{CSym}) to the characteristics obtained and classified
in \cite{AP1}, we get all the characteristic symmetries. Since the
Lagrange anchor is strongly integrable, characteristic symmetries
form an infinite dimensional Lie subalgebra in the Lie algebra of
all symmetries. This  subalgebra was previously unknown. For low
spins ($s=1/2, 1$) the Lie algebra of characteristic symmetries
contains a finite dimensional subalgebra which is isomorphic to the
Lie algebra of conformal group. The elements of this subalgebra
correspond to conserved currents that are expressible in terms of
the energy-momentum tensor.

\section*{Conclusion}

We have presented a Poincar\'e invariant Lagrange anchor
for the Bargmann-Wigner equations.  By making use  this Lagrange anchor we have established a systematic connection between the symmetries and conservation laws of the equations. The Lagrange anchor, being
independent of fields, is strongly integrable. As a consequence
the symmetries associated with the conservation laws (characteristic
symmetries) form an infinite-dimensional subalgebra in the full  Lie
algebra of symmetries. The physical meaning of this subalgebra remains unclear for us
 at the moment.

The Lagrange anchor (\ref{ULA}) may be used for quantization of the
Bargmann-Wigner equations. At the free level the corresponding
generalized Schwinger-Dyson equations and probability amplitude was
found in \cite{KLS5}. It can also be  a good starting point for
constructing the Lagrange anchor for  Vasiliev's equations and
development of a quantum theory of higher-spin interactions.

\subsection*{Acknowledgments}
We are thankful  to G. Barnich,  E.D. Skvortsov, and M.A. Vasiliev
for discussions on various topics related to this work and for
relevant references.

\end{document}